\documentclass[twocolumn,showpacs,preprintnumbers,amsmath,amssymb]{revtex4}
\usepackage{graphicx}% Include figure files
\usepackage{dcolumn}% Align table columns on decimal point
\usepackage{bm}% bold math
\usepackage{oldgerm}
\usepackage{amsmath}
\usepackage{psfrag}
\usepackage{color}
\usepackage{latexsym,amsfonts}
\usepackage{graphicx}
\newcommand{\bee}{\begin{eqnarray}}
\newcommand{\eee}{\end{eqnarray}}
\newcommand{\be}{\begin{equation}}
\newcommand{\ee}{\end{equation}}

\begin{document}
%\preprint{DRAFT}

\title{%The elastic electron-deuteron scattering beyond one-photon exchange. Polarization observables. 
Two-photon exchange and elastic scattering of longitudinally polarized electron on polarized deuteron}
\author{Alexander~P.~Kobushkin}
% \email{kobushkin@bitp.kiev.ua}
\affiliation{%
Bogolyubov Institute for Theoretical Physics, Metrologicheskaya Street. 14B\\ 
03680 Kiev, Ukraine
}%
\author{Yaroslav~D.~Krivenko-Emetov}
\affiliation{%
Institute for Nuclear Research, Prospect Nauki 47\\ 
03680 Kiev, Ukraine
}%
% \email{yakr@kinr.kiev.ua}%
\author{Stanislav~Dubni\v cka}
% \email{}
\affiliation{%
Institute of Physics, Slovak Academy of Sciences, Bratislava, Slovak Republic
}%
\author{Anna~Z.~Dubni\v ckova}
% \email{}
\affiliation{%
Department of Theoretical Physics, Comenius University, Bratislava, Slovak Republic
}%
%\\

\date{\today}% It is always \today, today,

\begin{abstract}
Structure functions and polarization observables in elastic scattering of longitudinally polarized electron on polarized deuteron are considered within approximation of one-photon~+~two-photon exchange. It is shown that contribution of two-photon exchange in the generalized structure function $\mathcal A$ is of order of few percent, while in the generalized structure function $\mathcal B$ it is of order of 10--20\%. We have found that components $T_{20}$ and $T_{21}$ of tensor analyzing power are mainly determined by one-photon exchange, but $T_{22}$ is mainly determined by interference between one-photon exchange and two-photon exchange. We have also considered polarization observables $T_{11}$, $C_{21}$ and $C_{22}$ which are proportional to imaginary part of the reaction amplitude and vanish in the framework of one-photon exchange.
\end{abstract}

% PACS, the Physics and Astr%onomy
\maketitle
\section{\label{sec:Introduction}Introduction}
Over resent years, the interest to study of polarization observables in electron scattering on hadron systems (nucleons, pions, lightest nuclei) has been raised. This interest is based on the significant progress in experimental technique, as well as on the fact that polarization measurements give an opportunity to get more information about structure of the hadron systems, than study of unpolarized cross section. For example, polarization data obtained in resent years together with unpolarized cross sections make it possible to extract important information about effects beyond Born approximation in electron-proton scattering (see, e.g., \cite{Atkinson} and references therein).

In case of electron-deuteron scattering polarization measurements play an important role already in one-photon exchange (OPE) approximation. Indeed, the deuteron, due to its spin structure, has three electromagnetic form factors (the charge, $G_C$, quadrupole, $G_Q$, and magnetic, $G_M$, form factors), which are real functions of one variable, $Q^2$. The Rosenbluth separation allows to perform  and obtain two structure functions of the deuteron: $A(Q^2)$ (a combination of $G^2_C$, $G^2_Q$ and $G^2_M$) and $B(Q^2)$ (proportional to $G^2_M$). To separate all three form factors we need to measure an additional observable. Usually this is $t_{20}$ component of tensor polarization of the deuteron.

Similarly to elastic electron-nucleon scattering, two-photon exchange (TPE) is one of the most important effects beyond Born approximation in elastic $ed$-scattering \cite{Lev,Beijing,Dong2010,KKD}. 

Following Ref.~\cite{KKD}, Feynman diagrams for TPE diagrams in $ed$ scattering fall into two types: diagrams, where both intermediate photons interact with the same nucleon, (type I) and diagrams, where photons interact with different nucleons, (type II). In Ref.~\cite{Lev} diagrams of type II were calculated using the simplest gaussian wave function of the deuteron. In turn, in Refs.~\cite{Beijing,Dong2010} some effects connected with diagrams of type I were examined in the framework of effective Lagrangian approach \cite{Ivanov,Dong}. Both types of diagrams were calculated simultaneously in Ref.~ \cite{KKD} within semi-relativistic approximation with the deuteron wave functions for ``realistic'' NN~potentials. 

% The aim of the present paper is to study how strongly TPE affects observables in the elastic scattering of longitudinally polarized ultra-relativistic electron on polarized deuteron.

The aim of the present paper is to study differential cross section and polarization observables in the elastic scattering of longitudinally polarized ultra-relativistic electron on polarized deuteron in the framework of OPE+TPE approximation.

The paper is organized as follows. In Sec.~\ref{sec:observables} we discuss density matrix for spin-1 particle and define polarization observables in elastic scattering of longitudinally polarized electron on polarized deuteron. In Sec.~\ref{sec:second_order} the polarization observables are calculated in the framework of OPE+TPE approximation. Numerical results and discussion are given in Sec.~\ref{sec:Numerical}.
\section{\label{sec:observables}Density matrix for spin-1 particle and observables}
\begin{figure}
\centering 
%\hspace{-2.cm}
\includegraphics[height=0.25\textheight]{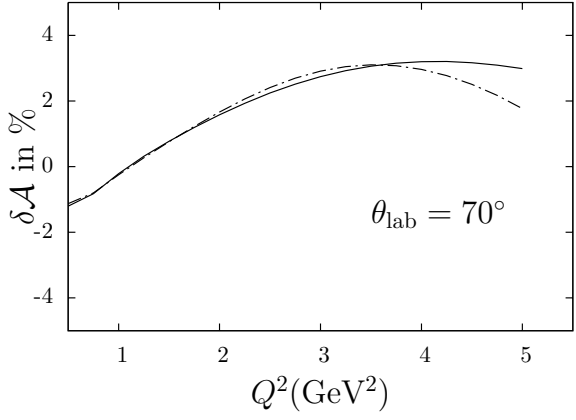}
\caption{TPE correction to the generalized structure function $\mathcal A$ at $\theta_\mathrm{lab}=70^\circ$. The solid and dash-doted lines are for CD-Bonn and Paris wave functions, respectively.}
\label{fig:A_and_B}
\end{figure} 
\begin{figure}
\centering 
%\hspace{-2.cm}
\includegraphics[height=0.325\textheight]{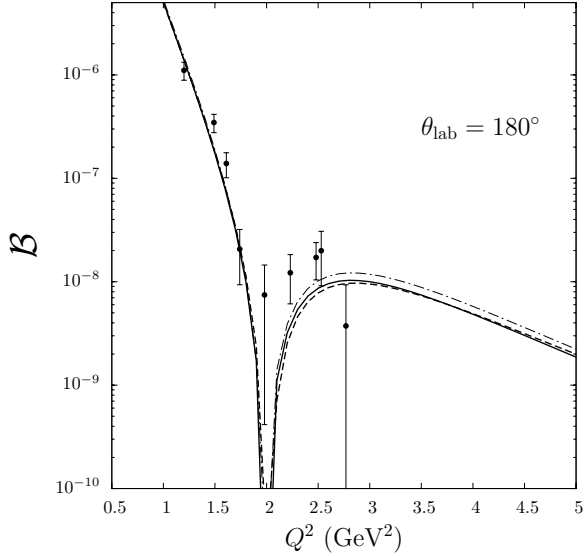}
\caption{The generalized structure function $\mathcal B$ at $\theta_\mathrm{lab}=180^\circ$. The solid and dashed lines are for OPE+TPE calculations with  CD-Bonn and Paris NN-potentials, respectively, dot-dashed is for OPE-parametrization. Data are from Ref.~\cite{bosted}.}
\label{fig:B180}
\end{figure} 
Later on we will work in spherical basis and use a right-hand coordinate system in which the positive $z$-axis is directed along transfer momentum $\vec q$ and $y$-axis is directed along $\vec k\times \vec k'$, where $k_\mu$ and $k_\mu'$ are momenta of incoming and outgoing electrons.

In this case, according to Madison Convention \cite{Madison}, the density matrix of a spin-1 particle (the deuteron) is given by the relation
\be\label{density_matrix1}
\rho=\tfrac13\sum_{kq}t^\ast_{kq}\tau_{kq},
\ee
where $t_{kq}$ are polarization parameters of the deuteron and $\tau_{kq}$ are spherical tensors. The later are expressed 
\begin{equation}\label{spherical_tensors}
\begin{split}
&\tau_{00}=1,\\
&\tau_{10}=\sqrt{3/2}S_z
\qquad
\tau_{1\pm 1}=\mp (1/2)\sqrt{3}(S_x\pm iS_y),
\\
&\tau_{2\pm 1}= \mp (1/2)\sqrt{3}[(S_x \pm iS_y)S_z+S_z(S_x \pm iS_y)],\\
&\tau_{2\pm 2}=(1/2)\sqrt{3}(S_x \pm iS_y)^2
 \end{split}%\label{tensors}
 \end{equation}
through the deuteron spin matrices $\vec S$
\be\label{Sx}
\begin{split}
&S_x=\frac{1}{\sqrt{2}}\left(
\begin{array}{ccc}
0 & 1 & 0\\
1 & 0 & 1\\
0 & 1 & 0
\end{array}
\right),
\quad
S_y=\frac{1}{\sqrt{2}}\left(
\begin{array}{ccc}
0 & -i & 0\\
i & 0 & -i\\
0 & i & 0
\end{array}
\right),\\
%\qquad
&S_z=\left(
\begin{array}{ccc}
1 & 0 & 0\\
0 & 0 & 0\\
0 & 0 & -1
\end{array}
\right)
\end{split}
\ee

From (\ref{spherical_tensors}) and hermiticity of the spin operator we immediately get
\be\label{spherical_tensors_condition}
\tau^\dag_{kq}=(-1)^q\tau_{k-q}
\ee
and the hermiticity condition for the density matrix yields for $t_{kq}$ 
\be\label{hermithity}
t^\ast_{kq}=(-1)^qt_{k-q}.
\ee
% %
% The deuteron density matrix is given by
% %
% \be\label{density_matrix2}
% \rho=\tfrac13\sum_{kq}(-1)^qt_{kq}\tau_{k-q}.
% \ee
% %
After that we come to explicit expression of the deuteron density matrix
\begin{widetext}
\be\label{density_matrix3}
\rho=\tfrac13\left( 
\begin{array}{ccc}
1+\sqrt{\tfrac32}t_{10} + \tfrac1{\sqrt2} t_{20}
&
\sqrt{\tfrac32}\left(t_{1-1} + t_{2-1}\right)
&
\sqrt{3}\, t_{2-2}
\\
-\sqrt{\tfrac32}\left(t_{11} + t_{21}\right)
&
1-\sqrt2\,t_{20}
&
\sqrt{\tfrac32}\left(t_{1-1} - t_{2-1}\right)
\\
\sqrt{3} \,t_{22}
&
-\sqrt{\tfrac32}\left(t_{11} - t_{21}\right)
&
1-\sqrt{\tfrac32}t_{10} + \tfrac1{\sqrt2}t_{20}
\end{array}
\right) 
\ee
\end{widetext}

The cross section for elastic scattering of electron with helicity sign $h$ on polarized deuteron is given by
\be\label{cross_section}
\frac{d\sigma(h)}{d\Omega}=\frac{d\sigma_0}{d\Omega}\left(1+\sum_{k=1}^2\sum_{q=-k}^{k}t_{kq} {T^{h}_{kq}}^\ast \right),
\ee
where $\frac{d\sigma_0}{d\Omega}$ is cross section for unpolarized particles and $t_{kq}$ and $T^{h}_{kq}$ are polarization tensor of the incoming deuteron and analyzing power of the reaction, respectively.

The analyzing power is given by
\begin{equation}\label{oservables}
T^h_{kq}=\frac{\mathrm{Tr} [\mathcal M^h\tau_{kq}{\mathcal M^h}^\dag]}{\tfrac12\sum_{h} \mathrm{Tr} [\mathcal M^h {\mathcal M^h}^\dag]},
\end{equation}
where $\mathcal M^h$ is scattering amplitude. Similarly to polarization tensor, $T^h_{kq}$ obeys the hermiticity condition
\be\label{hermiticity}
{T^h_{kq}}^\ast=(-1)^qT^h_{k-q}.
\ee
Furthermore, parity conservation implies
\be\label{parity}
T^h_{kq}=(-1)^{k+q}T^{-h}_{k-q}.
\ee
These two conditions allow us to write
\be\label{paramet}
\begin{split}
&T_{10}^h=h C_{10}^L,\\
&T_{11}^h=iT_{11}+h C_{11}^L,\qquad T_{1-1}^h=iT_{11}-h C_{11}^L,\\
&T_{20}^h=T_{20},\\
&T_{21}^h=T_{21}+ih C_{21}^L,\qquad T_{2-1}^h=-T_{21}+ih C_{21}^L,\\
&T_{22}^h=T_{22}+ih C_{22}^L,\qquad T_{2-2}^h=T_{22}-ih C_{22}^L,
\end{split}
\ee
where $T_{kq}$ and $C_{kq}^L$ are purely real. We call $T_{kq}$ and $C_{kq}^L$  analyzing power and correlation parameter, respectively. It should be noted that OPE amplitude is real and $T_{11}$, $C_{21}^L$ and $C_{22}^L$ are nonvanishing only beyond OPE approximation.

Substituting (\ref{paramet}) in (\ref{cross_section}) and using relation (\ref{hermithity}) we get the differential cross section for elastic scattering of electron with longitudinal polarization $p$ on polarized deuteron
\be\label{cross_section_pol}
\begin{split}
\frac{d\sigma}{d\Omega}=\frac{d\sigma_0}{d\Omega}&\left[
1+2\Im\mathrm m t_{11}T_{11}+t_{20}T_{20}+2\Re\mathrm e t_{21}T_{21}+\right.\\
&\left.+2\Re\mathrm e t_{22}T_{22}+p\left( t_{10}C_{10}^L +2\Re\mathrm e t_{11}C_{11}^L +\right.\right.\\
&\left.\left. +2\Im\mathrm m t_{21}C_{21}^L+ 2\Im\mathrm m t_{22}C_{22}^L\right) 
\right].
\end{split}
\ee
%
%where $p$ is longitudinal polarization of the electron.
%
\section{\label{sec:second_order}Polarization observables in the second order perturbation calculations}%
\begin{figure}
\centering 
\hspace{-1.5cm}
\includegraphics[height=0.25\textheight]{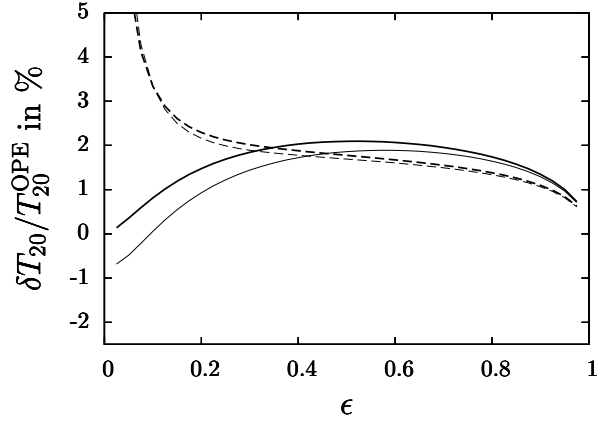}
\caption{The TPE correction to $T_{20}$ at $Q^2=$2 and 3 GeV$^2$ (solid and dashed respectively). The bold and thin lines are for calculations with CD-Bonn and Paris NN-potentials. $T_{20}^\mathrm{OPE}\approx$0 at $Q^2=1$~GeV$^2$.}
\label{fig:T20}
\end{figure} 
%\footnote{\footnote{}}
\begin{figure}
\centering 
\hspace{-1.5cm}
\includegraphics[height=0.25\textheight]{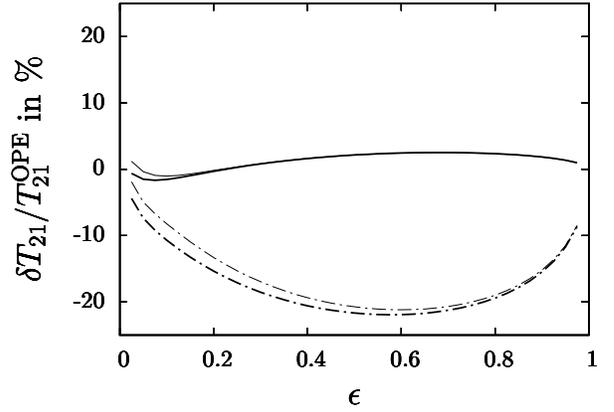}
\caption{The TPE correction to $T_{21}$ at $Q^2=$1 and 3 GeV$^2$ (solid and dot-dashed lines, respectively). The bold and thin lines are for calculations with CD-Bonn and Paris NN-potentials. $T_{21}^{\rm OPE}\approx$0 at $Q^2\approx$2 GeV$^2$ and relative correction becomes infinite.}
\label{fig:T21}
\end{figure} 
\begin{figure}
\centering 
%\hspace{-2.cm}
\includegraphics[height=0.25\textheight]{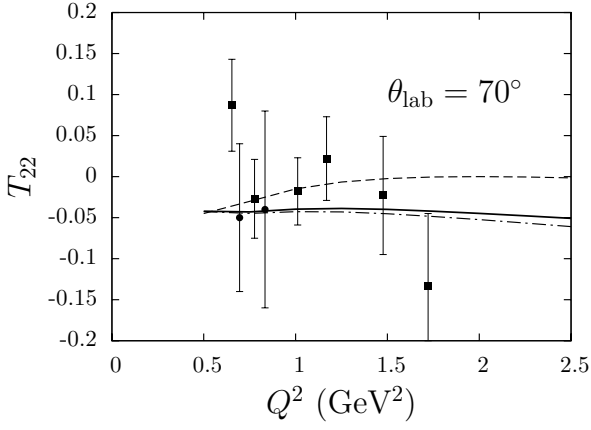}
\caption{$T_{22}$ at $\theta_\mathrm{lab}=70^\circ$. The dashed line is for OPE calculations, the solid and dash-doted lines are for OPE+TPE calculated with CD-Bonn and Paris NN-potentials, respectively. Data are from \cite{Garcon} and \cite{Abbott} (circles and boxes, respectively).}
\label{fig:T22}
\end{figure} 
\begin{figure}
\centering 
%\hspace{-1.5cm}
\includegraphics[height=0.25\textheight]{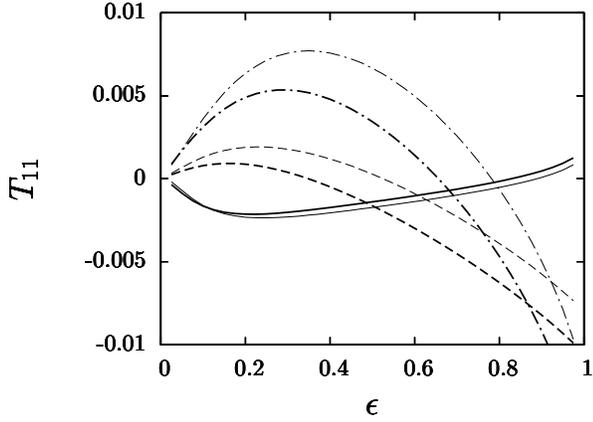}
\caption{$T_{11}$ at $Q^2=$1, 2 and 3 GeV$^2$ (solid, dashed and dot-dashed lines, respectively). The bold and thin lines are for calculations with CD-Bonn and Paris NN-potentials.}
\label{fig:T11}
\end{figure}
It follows from $P$ and $T$ invariance that elastic scattering amplitude of a spin $\frac12$-particle (electron) on a spin-1 particle (deuteron) is determined by 12 invariant amplitudes. Putting the electron mass to zero reduce the number of invariant amplitudes (form factors) to 6 (see, e.g., \cite{Dong2006,Gakh}).

All calculations will be done in the Breit frame. In this frame it is useful, instead of usual amplitude $\mathcal M$, to introduce reduced amplitude $\mathcal T_{\lambda'\lambda, h}$
\be \label{Gen.Amplitue}
\mathcal M=\frac{16\pi\alpha}{Q^2}E_eE_d \mathcal T_{\lambda'\lambda,h},
\ee
where $\alpha\approx 1/137$ is the fine-structure constant, $\lambda$, $\lambda'$ are spin projections of the incoming and outgoing deuteron and $E_e$ and $E_d$ are energies of the electron and deuteron in the Breit frame. For the reduced amplitude we will adopt the following parametrization \cite{KKD}
\be\label{Gen.Reduce_Amplitue}
\mathcal T_{\lambda'\lambda; h}=\left(
\begin{array}{ccc}
\mathcal G_{11}\cos\frac{\theta}2&-\sqrt\frac{\eta}{2}\mathcal G_{10}^{h} & \mathcal G_{1,-1}^{h}\\[0.25cm]
\sqrt\frac{\eta}{2}\mathcal G_{10}^{-h} &\mathcal G_{00}\cos\frac{\theta}2&-\sqrt\frac{\eta}{2}\mathcal G_{10}^{h}\\[0.25cm]
\mathcal G_{1,-1}^{-h} & \sqrt\frac{\eta}{2}\mathcal G_{10}^{-h}& \mathcal G_{11}\cos\frac{\theta}2
\end{array}
\right),
\ee
where $\theta$ is scattering angle in the Breit frame, $\eta=Q^2/(4M^2)$, $M$ is the deuteron mass and
\be \label{Gen.Gmn}
\mathcal G_{10}^{h}=f_1+h\sin\tfrac{\theta}2 f_2, \qquad
\mathcal G_{1,-1}^{h}=f_3+h\sin\tfrac{\theta}2 f_4.
\ee
The form factors $\mathcal G_{11}$, $\mathcal G_{00}$, $f_1$, ..., $f_4$ are complex functions of the two independent kinematical variables, for example, $Q^2$ and the commonly used polarization parameter
%
%\be\label{epsilon}
$\epsilon=\frac{\cos^2\frac{\theta}2}{1+\sin^2\frac{\theta}2}$.
%\ee
%

Instead of the the form factors  $\mathcal G_{11}$, $\mathcal G_{00}$, $f_1$, ..., $f_4$ the following linear combinations are introduced
\be \label{GcGqGm}
\begin{split}
&\mathcal G_C=\frac13\left(\mathcal G_{11}+2 \mathcal G_{00}\right),\qquad
\mathcal G_Q=\frac{1}{2\eta}\left( \mathcal G_{00}-\mathcal G_{11}\right),\\
&\mathcal G_M=\frac{f_1+\sin^2\tfrac{\theta}2f_2}{1+\sin^2\tfrac{\theta}2},
\qquad g_1=\frac{f_1-f_2}{1+\sin^2\tfrac{\theta}2},\\
&g_2=f_3,\qquad g_3=f_4.
\end{split}
\ee
%
% %
% \be \label{GcGqGm}
% \begin{split}
% &\mathcal G_{11}=\mathcal G_C-\tfrac23\eta \mathcal G_Q,\qquad
% \mathcal G_{00}=\mathcal G_C+\tfrac43\eta \mathcal G_Q,\\
% &f_1=\mathcal G_M + g_1\sin^2\tfrac{\theta}2,\qquad f_2=\mathcal G_M - g_1,\\
% &f_3=g_2,\qquad f_4=g_3.
% \end{split}
% \ee
% %
We call $\mathcal G_C(Q^2,\epsilon)$, $\mathcal G_Q(Q^2,\epsilon)$ and $\mathcal G_M(Q^2,\epsilon)$ the generalized electric, quadrupole and magnetic from factors. In zeroth order in $\alpha$ the  generalized electric, quadrupole and magnetic from factors are reduced to the usual electric, quadrupole and magnetic form factors, $G_C(Q^2)$, $G_Q(Q^2)$ and $G_M(Q^2)$, while the form factors $g_1(Q^2,\epsilon)$, $g_2(Q^2,\epsilon)$ and $g_3(Q^2,\epsilon)$ are of order $\alpha$ and vanish in the Born approximation.

Substituting  (\ref{Gen.Reduce_Amplitue})-(\ref{GcGqGm}) in Eq.~(\ref{oservables}) we arrive at the following expressions for analyzing powers
\be\label{analizing_powers}
\begin{split}
 T_{20}=&\dfrac{-\eta}{3\sqrt{2}\mathcal S}\left[8\left( \Re \mathrm e\left( \mathcal G_C^\ast \mathcal G_Q\right) +\frac{\eta}{3}|\mathcal G_Q|^2\right) +\right.\\
&\left.+(1+2\mathrm{tg}^2\tfrac{\theta}2)|\mathcal G_M|^2 \right],\\
T_{21}=&-\sqrt{\frac{\eta }{3}}\frac{1}{\cos^2\tfrac{\theta }{2}\mathcal S} \cdot \left[
   2 \cos\tfrac{\theta }{2}\eta \Re \mathrm e \left( \mathcal G_M^\ast  \mathcal G_Q\right) 
  +\right.\\
&\left.+G_M\Re \mathrm e\left(\sin ^2\tfrac{\theta}{2} g_3 + g_2\right)
  +\right.\\
&\left.+2 \sin ^2\tfrac{\theta }{2}\cos\tfrac{\theta }{2}\eta \Re\mathrm e g_1 G_Q \right],\\
T_{22}=&\frac{1}{2\sqrt{3}\cos^2\tfrac{\theta }{2}\mathcal S}\left[-\cos^2\tfrac{\theta }{2} \eta  |\mathcal G_M|^2 
   -\right.\\
&\left.-4 \sin ^2\tfrac{\theta }{2}\eta G_M \Re \mathrm e g_1 
   +\right.\\
&\left.+4\cos \tfrac{\theta }{2} (G_C -\tfrac23\eta G_Q)\Re\mathrm e g_2\right],\\
T_{11}=&
   \frac{\sqrt{\eta }}{3\sqrt{3}\cos^2\tfrac{\theta }{2}\mathcal S} \times\\
&\times\left[2 \cos\tfrac{\theta }{2}\Im \mathrm m \left( \mathcal G_M^\ast \left(3 \mathcal G_C+\eta \mathcal G_Q\right)\right)-
   \right.\\
&\left.
  -2\sin ^2\tfrac{\theta}{2} \cos \tfrac{\theta}{2} \left(3 G_C+\eta  G_Q\right)\Im \mathrm m g_1
  -\right.\\
&\left.-3 G_M\Im\mathrm m\left(\sin ^2\tfrac{\theta}{2}g_3 +g_2\right)\right],
\end{split}
\ee
and polarization correlations
\be\begin{split}
C_{21}^L=&\sqrt{\frac{\eta }{3}} \frac1{\cos^2\tfrac{\theta }{2}\mathcal S}\times\\
&\times\left[
   -\eta  \sin \theta \left(\Im\mathrm m\left( \mathcal G_M^\ast\mathcal G_Q\right) 
   + G_Q\Im\mathrm m g_1 \right)
   +\right.\\
&\left.+\sin \tfrac{\theta }{2}G_M\Im\mathrm m\left(g_2+g_3\right) \right],\\
C_{22}^L=&-\frac{1}{\sqrt{3}\cos^2\tfrac{\theta }{2}\mathcal S}\times\\
&\times\left[
\sin\theta(G_C-\tfrac23\eta G_Q) \Im\mathrm m g_3 +\right.\\
&\left.+\sin\tfrac{\theta }{2} \left(\sin ^2\tfrac{\theta }{2}+1\right)\eta G_M \Im\mathrm m g_1\right],\\
C_{10}^L=&
\sqrt{\frac{2}{3}}\eta\mathrm{tg} \tfrac{\theta}{2} \cdot \frac{  
   -|\mathcal G_M|^2+\cos ^2\frac{\theta}{2}G_M\Re\mathrm e g_1 }
{\cos \frac{\theta}{2}\mathcal S},\\
C_{11}^L=&
\frac{\sqrt{\eta }}{3\sqrt{3}\cos^2\tfrac{\theta}{2}\mathcal S} \left\{\sin \theta \left[
\Re \mathrm e \left( \mathcal G_M^\ast (3 \mathcal G_C+\eta\mathcal G_Q)\right) -\right.\right.\\
&\left.\left. - \left(3G_C+\eta  G_Q\right)\Re\mathrm e g_1\right] +\right.\\
&\left.+ 3\sin\tfrac{\theta}{2} G_M\Re\mathrm e\left(g_2+g_3\right)\right\}, 
\end{split}
\ee
where we neglected terms of order $\alpha^2$ and thus
$\mathcal G_K^\ast \mathcal G_L=G_K G_L + \delta\mathcal G_K^\ast G_L+ G_K\delta\mathcal G_L$, $(K,L)=(C,Q,M)$, $\mathcal G_K=G_K+\delta \mathcal G_K$, $\mathcal G_L=G_L+\delta \mathcal G_L$;
\be\label{S}
\begin{split}
& 
\mathcal S=\mathcal A+\mathcal B\mathrm{tg}^2\tfrac{\theta_{\mathrm{lab}}}2,
\end{split}
\ee
with
\be\label{AB}
\begin{split}
& \mathcal A(Q^2,\theta)=|\mathcal G_C|^2+\tfrac89\eta^2|\mathcal G_Q|^2 +\tfrac23\eta|\mathcal G_M|^2,\\
& \mathcal B(Q^2,\theta)=\tfrac43 (1+\eta)\eta|\mathcal G_M|^2
\end{split}
\ee
are generalized structure functions and $\mathrm{tg}^2\theta_{\mathrm{lab}}/2=(1+\eta)^{-1}\mathrm{tg}^2\theta/2$.
\begin{figure}
\centering 
%\hspace{-2.cm}
\includegraphics[height=0.25\textheight]{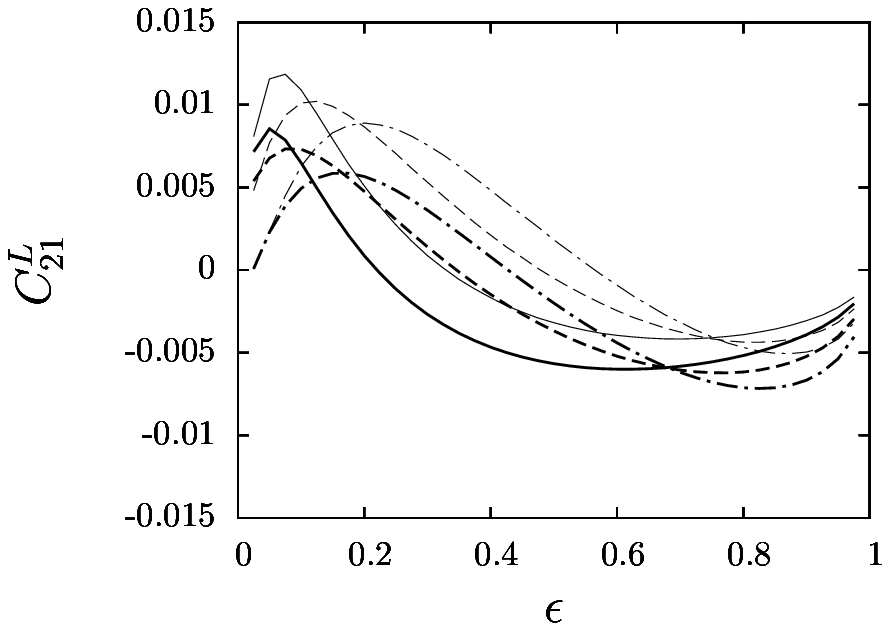}\\
%\hspace{-1.5cm}
\includegraphics[height=0.25\textheight]{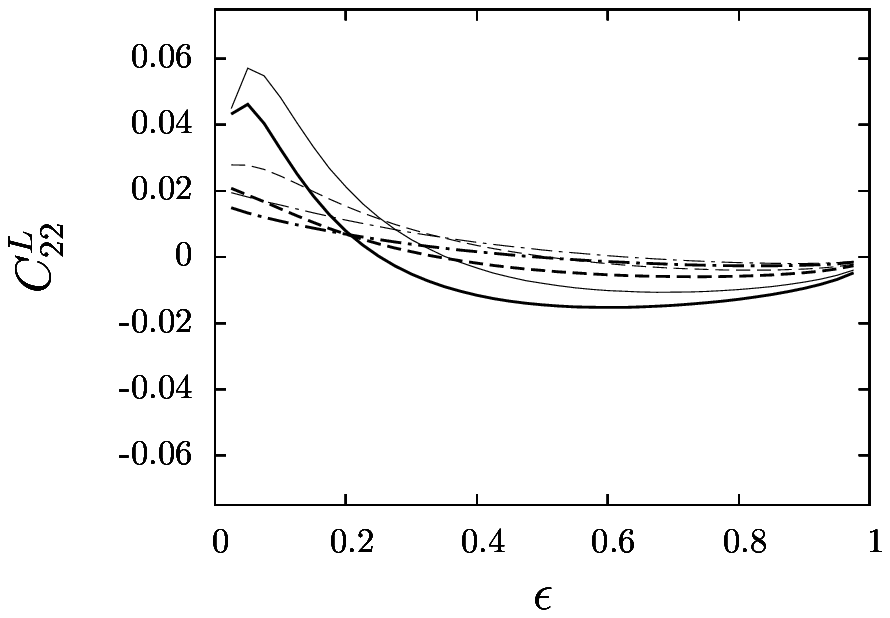}
\caption{$C_{21}^L$ and $C_{22}^L$ at $Q^2=$1, 2 and 3 GeV$^2$ (solid, dashed and dot-dashed lines, respectively). The bold and thin lines are for CD-Bonn and Paris NN-potentials.}
\label{fig:C21}
\end{figure}
\section{\label{sec:Numerical}Numerical results and conclusions}
\begin{figure}
\centering 
%\hspace{-2.cm}
\includegraphics[height=0.25\textheight]{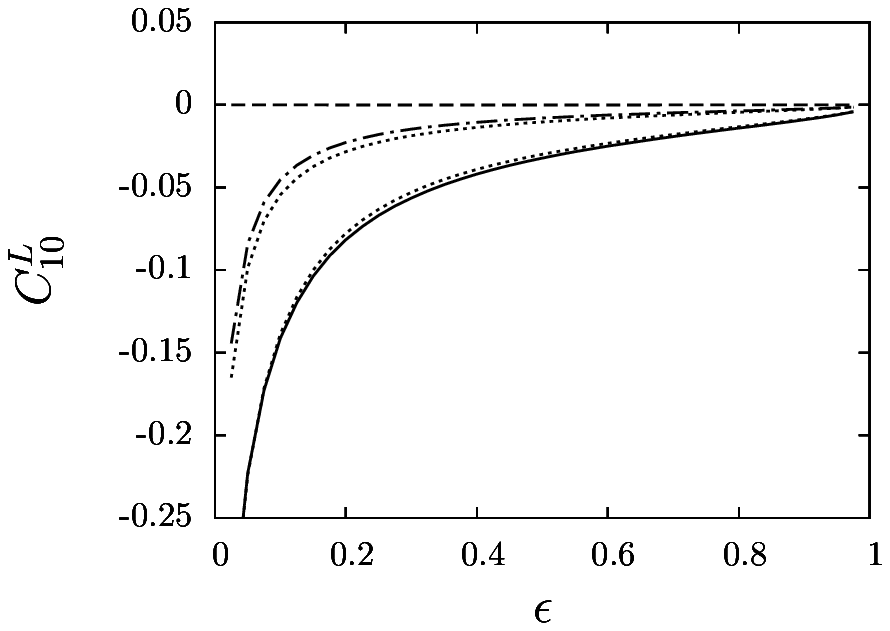}\\
%\hspace{-1.5cm}
\includegraphics[height=0.25\textheight]{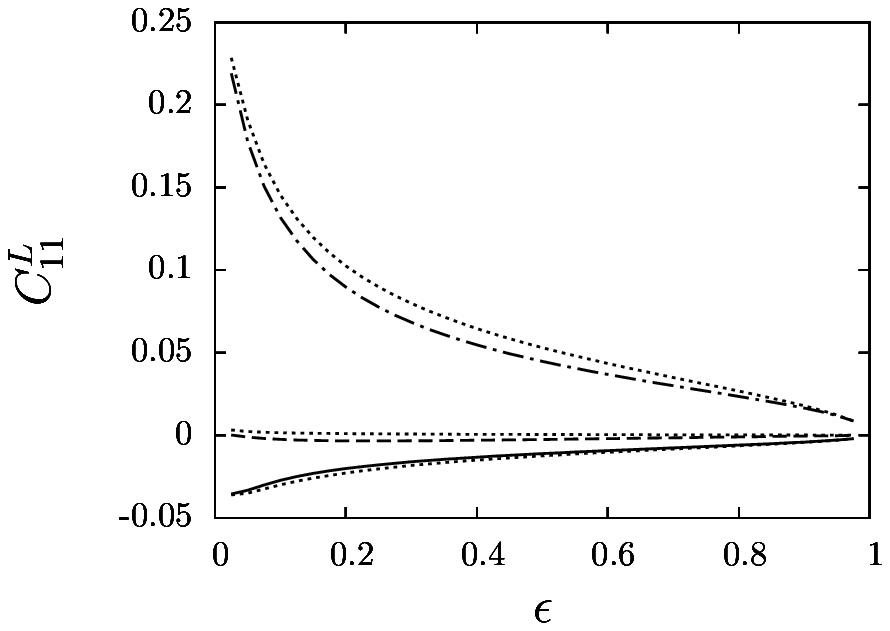}
\caption{$C_{22}^L$ and $C_{10}^L$ at $Q^2=$1, 2 and 3 GeV$^2$ (solid, dashed and dot-dashed lines, respectively). The TPE calculations are for CD-Bonn potential. Dotted lines are for the OPE approximation.}
\label{fig:C11} 
%}
\end{figure}
In previous section the generalized structure functions $\mathcal A(Q^2,\theta)$ and $\mathcal B(Q^2,\theta)$ and polarization observables were expressed by generalized form factors. Now we will calculate behavior of all this quantities. 

In numerical calculations of the TPE amplitudes we used semi-relativistic approach of Ref.~\cite{KKD} with the deuteron wave function for CD-Bonn \cite{CD-Bonn} and Paris \cite{Paris} NN-potentials. The deuteron form factors $G_C(Q^2)$, $G_Q(Q^2)$ and $G_M(Q^2)$ were taken from parametrization of Ref.~\cite{KS} with parameters taken from fit of Ref.~\cite{Ball}.

We find that TPE contribution is of order of few percent (see Fig.~\ref{fig:A_and_B}) in the generalized structure function $\mathcal A$, while in the generalized structure function $\mathcal B$ it is more significant. For example, at $\theta=180^\circ$ and  $Q^2>2.5$~GeV$^2$ TPE effect are estimated to be 10 to 20~\%  in $\mathcal B$ (Fig.~\ref{fig:B180}). One also sees, that value of TPE correction is strongly dependent on the deuteron wave function. 

The $\epsilon$ dependence of relative TPE correction to $T_{20}$ and $T_{21}$ at few values of $Q^2$ are shown in Figs.~\ref{fig:T20} and \ref{fig:T21}. One can conclude that TPE is not significant in $T_{20}$. In $T_{21}$ the role of TPE effects increases with $Q^2$. For example, at $Q^2=$3 GeV$^2$ the TPE correction comes up to 20\%. In $T_{22}$ the interference between OPE and TPE amplitudes becomes a dominant contribution, Fig.~\ref{fig:T22}. The latter conclusion is obvious, because $G_C/G_M\sim$10 and the term $\frac83 \cos\frac{\theta} 2G_C \Re\mathrm e g_2$ ranks over the term $\cos^2\frac{\theta}2 G_M^2$ in the expression for $T_{22}$ (see Eqs.~(\ref{analizing_powers})).

Our estimations for the observables, which vanish in the OPE approximation ($T_{11}$, $C_{21}$ and $C_{22}$), are displayed in Figs.~\ref{fig:T11} and \ref{fig:C21}.

We have also found that TPE makes only slight changes (not more than few percent) in $C_{10}$ and $C_{11}$ correlations, Fig.~\ref{fig:C11}.

In conclusion, we have considered general structure of differential cross section for elastic scattering of longitudinally polarized electron on polarized deuteron. 

The generalized structure functions $\mathcal A$ and $\mathcal B$ and polarization observables in this reaction are calculated in the framework of the one-photon~+~two-photon exchange. We find that TPE contribution in the generalized structure function $\mathcal A$ is less than experimental errors (about few percent) and about 10-20\% in the generalized structure function  $\mathcal B$ at $Q^2>$2.5~GeV$^2$. 

While in $T_{20}$ TPE correction is minor, interference between OPE and TPE becomes dominant contribution in $T_{22}$ at $Q^2>$0.5~GeV$^2$. This means that $T_{22}$ may be a good object for experimental study of two-photon exchange in $ed$-scattering. We have also calculated polarization observables $T_{11}$, $C_{21}$ and $C_{22}$ which are proportional to the imaginary part of the reaction amplitude and vanish in the framework of one-photon exchange.

\section*{Acknowledgment} A.P.K. acknowledge the partial support of the Program `Fundamental Properties of Physical Systems under Extreme Conditions' launched by the Section of Physics and Astronomy of National Academy of Sciences of Ukraine. S.D. and A.Z.D. would like to thank the Slovak Grant Agency for Sciences VEGA for support under Grant No. 2/0009/10. We thank for D.L.~Borisyuk and E.A.~Strokovsly for reading manuscript and their comments.

%%%%%%%%%%%%%%%%%%%%%%%%%%%%%%%%%%%%%%%%%%%%%%%%


\begin{thebibliography}{99}
\bibitem{Atkinson}
J.~Arrington, P.G.~Blunden, and W.~Melnitchouk, arXiv:1105.0951 (to be published in Prog. Part. Nucl. Phys.).
\bibitem{Lev}
F.M.~Lev, Yad. Fiz., {\bf 21}, 89 (1975).
\bibitem{Beijing}
Yu Bing Dong and D.Y.~Chen, Phys. Lett., {\bf B 675}, 426 (2009).
\bibitem{Dong2010}
Yu Bing Dong, Phys. Rev., C {\bf 82}, 068202 (2010).
\bibitem{KKD}
A.P.~Kobushkin, Ya.D.~Krivenko-Emetov, and S.~Dubni\v cka, Phys. Rev., C {\bf 81}, 054001 (2010).
\bibitem{Ivanov}
A.N.~Ivanov, N.I.~Troitskaya, M.~Faber, and H.~Oberhummer, Phys. Lett., {\bf B 361}, 74 (1995); Z. Phys. A {\bf 358}, 81 (1997); Nucl. Phys. A {\bf 617}, 414 (1997).
\bibitem{Dong}
Y.~Dong, A.~Faessler, T.~Gutsche, and V.F.~Lubovitskij, Phys. Rev. C {\bf 78}, 035205 (2008).
\bibitem{Madison}
The Madison Convention. Proc. of the 3-d International Symposium on Polarization Phenomena in Nuclear Physics, Madison, 1970, ed. by H.H.~Berschall and W.~Haeberli, University of Wisconsin Press, Madison, WI, 1971, p. xxv.
\bibitem{Dong2006}
Y.B.~Dong, C.W.~Kao, S.N.~Yang, and Y.C.~Chen, Phys. Rev., C {\bf 74}, 064006 (2006).
\bibitem{Gakh}
G.I.~Gakh and E.~Tomasi-Gustaffson, Nucl. Phys. A {\bf 799}, 127 (2008).
\bibitem{CD-Bonn}
R.~Machleidt, Phys. Rev. C{\bf 63}, 024002 (2001).
\bibitem{Paris}
M.~Lacombe et al., Phys. Lett. B{\bf 101}, 139 (1981).
\bibitem{KS}
A.P.~Kobushkin and A.I.~Syamtomov, Phys. At. Nucl., {\bf 58}, 1477 (1995).
\bibitem{Ball}
D.~Abbott et al., Eur. Phys. J., A{\bf 7}, 421 (2000). 
\bibitem{bosted}
P.E.~Bosted et al., Phys. Rev., C{\bf 42}, 38 (1990).
\bibitem{Garcon}
M.~Garcon et al., Phys. Rev. Lett., {\bf 65}, 1733 (1990). 
\bibitem{Abbott}
D.~Abbott et al., Phys. Rev. Lett., {\bf 84}, 5053 (2000). 
\end{thebibliography}
\end{document}